# Study of Asymmetric Magnetization Reversal and Exchange Bias in FePt(L1$_0$)/FeCo/CoO/FeCo Magnetic Multilayer

**Sadhana Singh[1,2], V.R. Reddy[1], Dileep Kumar[1],***

[1]*UGC-DAE Consortium for Scientific Research, Khandwa Road, Indore-452017, India*
[2]*Nanomagnetism and Microscopy Laboratory, Department of Physics, Indian Institute of Technology Hyderabad, Kandi, Sangareddy, Telangana, 502285, India*

[*]*Corresponding Author Email: dkumar@csr.res.in; dileep.esrf@gmail.com*

**Abstract:** The effect of the saturation field on the magnetization reversal of FePt(L1$_0$)/FeCo/CoO/FeCo multilayer (ML) has been investigated to understand the origin of asymmetric magnetization reversal and its correlation with exchange bias (EB). In the ML structure, the bottom FeCo layer is coupled to the hard FePt(L1$_0$) layer, and the top FeCo layer is comparatively free due to the relatively more distance from it. The ML has been deposited under UHV conditions and characterized at each stage of growth using magneto-optical Kerr effect and x-ray reflectivity techniques. Magnetization reversal is further studied through domain imaging using the Kerr microscopy technique. The experimental findings reveal that ML exhibits asymmetrical magnetization reversal for a certain range of azimuthal angles for both 1.5kOe and 50kOe saturation fields; however, this angular range of asymmetry decreases with the increase in the saturation field. Furthermore, EB was absent at the low saturation field, whereas, EB, in addition to asymmetry, is observed at the large saturation field. The origin of asymmetry is attributed to non-collinearity between magnetic anisotropy axes of both FeCo layers. It results from the proximity effect through short-range Heisenberg exchange interaction via the CoO barrier layer. On the other hand, EB arises due to unidirectional anisotropy induced in the FePt layer due to the high saturation field. It is further proposed that asymmetry would disappear when unidirectional anisotropy is strong enough to align both the FeCo layers in the saturation direction leading to loss of the non-collinearity between them.

# 1. Introduction

Asymmetrical magnetization reversal refers to reversal through two different methods in each branch of a hysteresis loop [1,2,3,4,5]. In one branch of the hysteresis loop, magnetization reversal occurs through coherent domain wall nucleation and propagation, whereas in another, it occurs through domain wall rotation. This behavior is often described as an unusual feature of exchange coupled antiferromagnetic (AFM)/ferromagnetic (FM) system [1,2,3,4,5]. It is in contrast to that of FM materials that exhibit symmetric hysteresis with respect to the magnetic field axis.

Until now, extensive work has been done experimentally and theoretically to understand the origin of this asymmetric magnetization reversal [1,2,3,4,5]. The origin of asymmetry in exchange bias (EB) systems was initially correlated to the irreversibility of pinned uncompensated spins due to the training effect [2,5], where an asymmetry was observed during the first cycle of magnetization reversal and it disappears with subsequent cycling. According to J. McCord *et al.* [1], the training associated asymmetry is due to dispersion in anisotropies of the AFM and FM layers. Later on, few studies correlated asymmetry with the existence of the higher order anisotropies [1,6]. For example, according to M. R. Fitzsimmons *et al.* [6], the magnetization reversal mechanism depends upon the orientation of the cooling field with respect to the twinned microstructure of the AFM and on the descending/ascending branch of measured hysteresis with respect to the saturation field direction. However, this theory of the existence of twinned crystal structure as a necessary condition for asymmetry was rejected by Beckmann *et al.* [7] and J. Camareo *et al.* [2], who proposed that asymmetry being an intrinsic property of EB system, cannot be restricted to only higher order anisotropy systems. Asymmetry in an exchange coupled system is further found to be dictated by the competition between unidirectional exchange anisotropy ($K_E$) at the AFM/FM interface and uniaxial anisotropy ($K_U$) of the FM layer in terms of their relative orientation (collinear and non-collinear) [1,2,4,7, 8] and strength [2,9] with respect to each other and to the applied magnetic field ($H_{APP}$) [10].

However, it is to be noted that though asymmetry is usually associated with EB, it is not always the case. There are various AFM/FM systems which exhibit EB but not asymmetry [11,12,13,14,15,16,17]. Furthermore, asymmetric magnetization reversal has also been observed in systems other than the EB systems. For example, both in magnetoresistance and domain reversal behavior, asymmetry was observed in the interlayer coupled Fe/Cr/Fe system [18]. M. Gottwald *et al.* [19] also reported asymmetry in magnetization reversal of soft $Co_{88}Tb_{12}$ layer in dipolar coupled $Co_{74}Tb_{26}/Cu/Co_{88}Tb_{12}$ spin valve structure with perpendicular magnetic anisotropy. Asymmetric magnetization reversal was observed when the hard $Co_{74}Tb_{26}$ layer was not fully saturated and interpreted in terms of dipolar interaction between the domains of hard and soft FM layers. Dipolar coupling between two FM/AFM nanostructures can also give rise to asymmetry [20]. Hence, the origin of asymmetrical magnetization reversal and its correlation with EB is still not understood and requires further study.

With the above objective, in the present work, we studied the effect of varying strength of saturation field ($H_{SAT}$) on asymmetric magnetization reversal in the exchange coupled FePt($L1_0$)/FeCo/CoO/FeCo multilayer (ML). $H_{SAT}$ of the hard FePt($L1_0$) layer was varied to tune the coupling of the FePt($L1_0$) layer with two FeCo layers. The bottom FeCo layer is directly coupled with FePt($L1_0$) hard magnetic layer through exchange interaction, whereas the top FeCo layer is comparatively free to rotate. Magneto-optical Kerr effect (MOKE) for hysteresis loop measurement and Kerr microscopy for domain imaging have been used as magnetic characterization tools. The effect of non-collinear interlayer coupling due to the proximity effect and its competition with unidirectional anisotropy ($K_{UNI}$) has been discussed to explain the angular-dependent asymmetric magnetization reversal and EB in the ML systems.

## 2. Experimental

For FePt($L1_0$)/FeCo/CoO/FeCo ML, the FePt layer was initially deposited on Si substrate by co-sputtering Fe and Pt targets at a base pressure of $2 \times 10^{-6}$ mbar. FePt film was then transferred to a UHV chamber, where the surface of the FePt layer was cleaned using the ion beam sputtering technique. The film was then annealed at 823 K for 30 min to obtain the $L1_0$ ordered hard magnetic phase. FeCo and Co layers were then deposited on FePt($L1_0$) layer using the e-beam evaporation technique. Before deposition of the top FeCo layer, the Co layer was oxidized by flowing oxygen in the chamber and annealing the sample at 573 K to form the CoO layer.

The ML was characterized at each deposition stage using in-situ MOKE and x-ray reflectivity (XRR) to study its magnetic and structural properties. Ex-situ MOKE measurements were then performed as a function of azimuthal angle by varying remanence state of FePt($L1_0$) layer using different $H_{SAT}$ strengths. Furthermore, domain imaging was performed using Kerr microscopy to study the domain dynamics. Hysteresis behavior of the demagnetized ML by applying 50 kOe field normal to the film's surface was also studied.

# 3. Result and discussion:

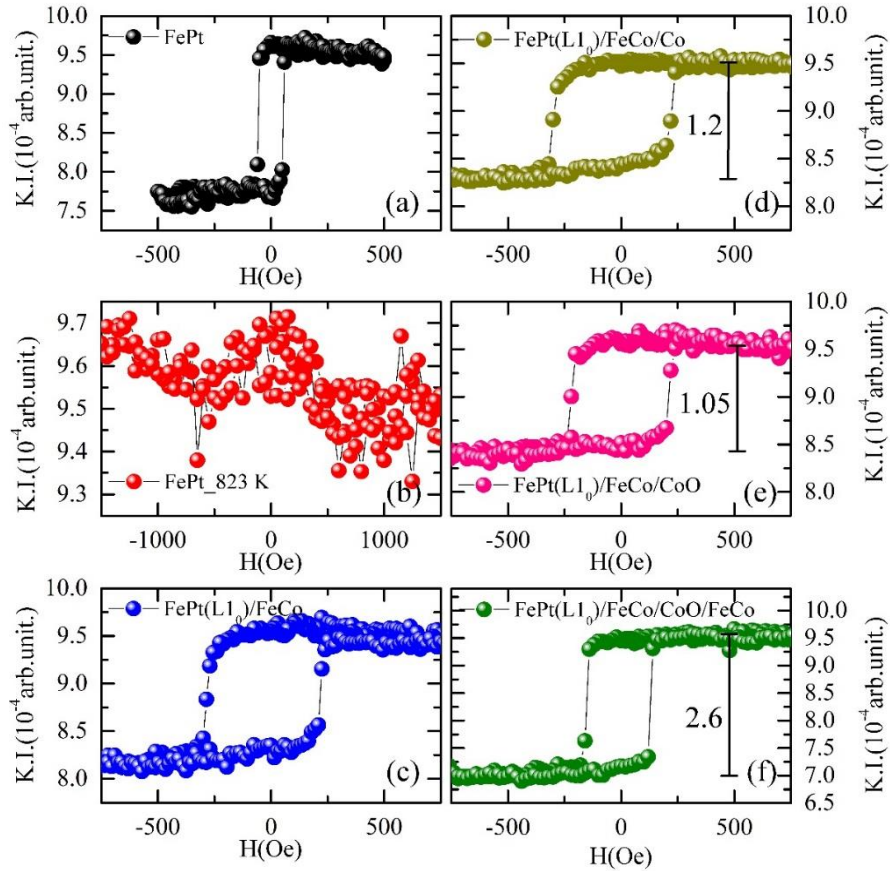

**Fig. 1:** In-situ MOKE hysteresis loops of (a) FePt in as-prepared stage, (b) FePt after annealing at 823 K (c) FePt(L1$_0$)/FeCo bilayer, (d) FePt(L1$_0$)/FeCo/Co trilayer (e) FePt(L1$_0$)/FeCo/CoO and (f) FePt(L1$_0$)/FeCo/CoO/FeCo ML at room temperature (RT).

Figure 1(a-f) show hysteresis loops of FePt(L1$_0$)/FeCo/CoO/FeCo ML obtained in-situ at various stages of growth. It was observed that the as-deposited FePt layer, which was used as a template for the ML structure, was soft FM in nature due to its disordered face-centered cubic structure [21]. After annealing at 823 K, no hysteresis loop was observed in the H$_{APP}$ range because the 1500 Oe field was insufficient to saturate the film due to the development of the hard magnetic L1$_0$ ordered face-centered tetragonal phase. Hysteresis loop of FePt(L1$_0$)/FeCo bilayer and FePt(L1$_0$)/ FeCo/Co exhibits weak EB. EB is attributed to exchange coupling between the hard FePt(L1$_0$) layer and soft bottom FeCo layer. However, EB is weak since moments of the FePt(L1$_0$) layer are randomly oriented and not aligned in a certain direction.

On the formation of the oxide layer on the bottom FeCo surface, it was observed that both Kerr signal and coercivity (H$_C$) decreased. The decrease in Kerr signal was due to oxidation of the Co layer resulting in the decrease in the net magnetic component of the film. The decrease in H$_C$ may be due to thermal annealing that removes defects and stress from the film. On the deposition of the top FeCo layer, the Kerr signal increases drastically, and no EB was observed. Note that due to the limited penetration depth of the laser [22], the Kerr signal of the complete ML in fig. 1f and the respective hysteresis loops here onwards in the manuscript corresponds to magnetization reversal of only the top FeCo layer. The absence of EB may be due to the weak magnetostatic interaction between FePt($L1_0$) and the top FeCo layer.

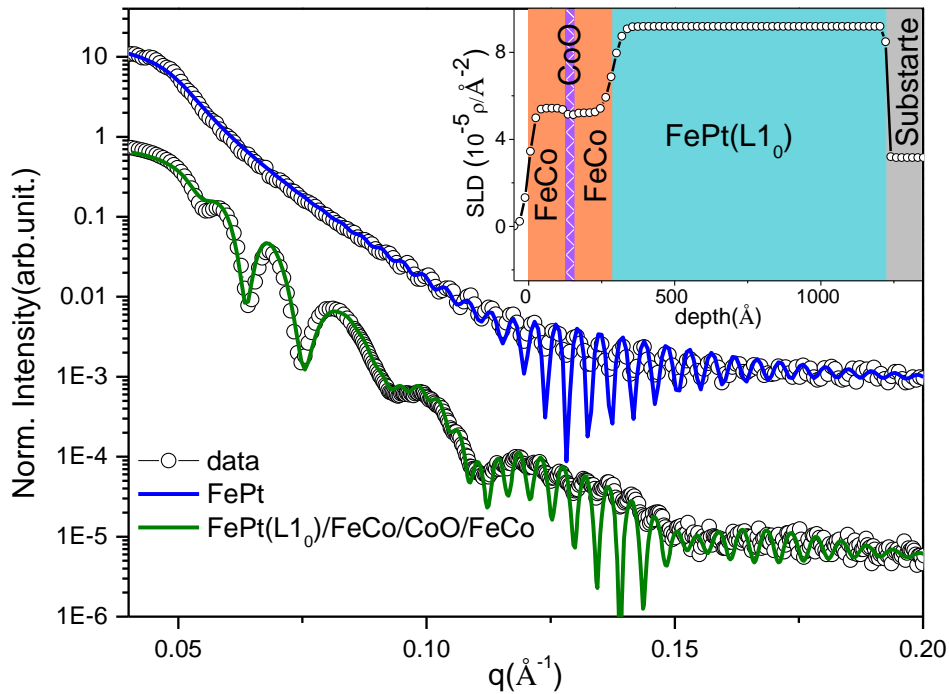

**Fig. 2:** XRR of (a) FePt template and (b) FePt($L1_0$)/FeCo/CoO/FeCo ML fitted using Parratt's formalism [23]. Inset shows SLD profile of FePt($L1_0$)/FeCo/CoO/FeCo sample obtained after fitting.

**Table 1.** Fitting parameters of XRR obtained using Parratt's formalism [23]. Error in layer thickness, as well as roughness, is ±0.5 Å.

| Layer | Thickness, d (Å) | Roughness, $\sigma$ (Å) | SLD, $\rho$ (Å$^{-2}$) |
| --- | --- | --- | --- |
| FeCo | 126.2 | 17.7 | 5.427E-5 |
| CoO | 30.4 | 9.8 | 5.081E-5 |
| FeCo | 131.2 | 10.6 | 5.212E-5 |
| FePt($L1_0$) | 940.0 | 26.0 | 9.184E-5 |

Figure 2 shows XRR patterns of the FePt template and FePt($L1_0$)/FeCo/CoO/FeCo ML fitted using Parratt's formalism [23]. Fitting parameters corresponding to the best fit are tabulated in table 1. Scattering length density (SLD) profile of the film obtained after fitting was also plotted in the inset of the figure. Fitting clearly shows the formation of a 30.4 Å CoO barrier layer between two FeCo layers.

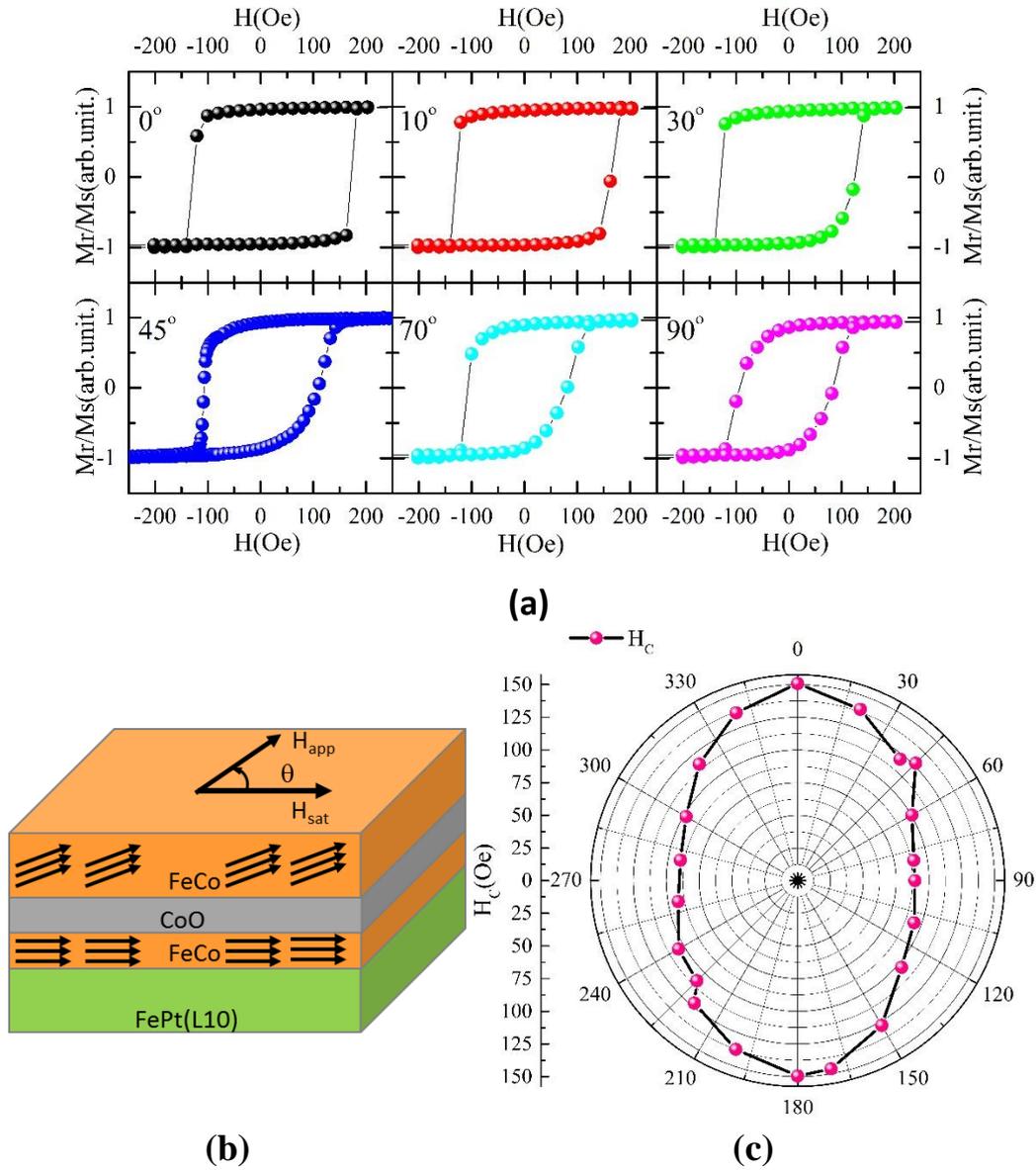

**Fig. 3:** (a) Hysteresis loop of FePt(L1$_0$)/FeCo/CoO/FeCo ML measured at various angles after applying H$_{SAT}$ =1.5 kOe . (b) Schematic diagram showing direction of H$_{SAT}$ and H$_{APP}$. (c) Polar plot of variation of H$_C$ as a function of azimuthal angle θ i.e. angle between H$_{SAT}$ and H$_{APP}$.

Figure 3(a) shows hysteresis loop of FePt(L1$_0$)/FeCo/CoO/FeCo ML as a function of azimuthal angle (θ) after applying 1.5 kOe magnetic field (H$_{SAT}$). Here, θ refers to the direction of H$_{APP}$ with respect to the H$_{SAT}$ (θ = 0° ). Significantly different behavior of hysteresis loops was observed at different θ values. A symmetric square hysteresis is observed along θ = 0°, indicating that magnetization reversal occurs through domain wall nucleation and propagation. Along θ = 90°, symmetrically rounded hysteresis loop with magnetic reversal through domain wall rotation was obtained. This behavior is in accordance with Stoner Wohlfarth's model for magnetization reversal of ferromagnetic films [24]. However, a clear asymmetry in the reversal pathway of two branches of the hysteresis loops was observed for a range of measuring angles between θ = 0° and θ = 90° with maxima along θ = 45°. Here, on one side of the hysteresis loop, the reversal occurs by domain wall nucleation and propagation, whereas on the other side, it occurs through domain wall rotation [1,2,4]. It is to be further noted that asymmetry was observed at angles away from the direction of $H_{SAT}$. Interestingly, in contrast to the previous studies [1,2,3,4,5,6], asymmetric magnetization reversal in the present case was observed even in the absence of EB. It indicates that $K_{UNI}$ may not play a role in the asymmetry. Asymmetry may be attributed to non-collinearity between the spins of dipolar coupled FeCo layer with respect to each other [25,26,27].

Domain images corresponding to magnetization reversal at different azimuthal angles (0°, 45°, and 90°) obtained using Kerr microscopy measurements are shown in fig. 4. It further confirms the fast reversal of domains with domain wall nucleation and propagation for θ = 0° (Fig. 4(a)). For θ = 90° (Fig. 4(c)), ripple-like images on both branches confirm magnetization reversal through domain wall rotation. However, for θ = 45° (Fig. 4(b)), the image at point F shows nucleation of a 180° domain that grows with the increasing magnetic field during reversal. In contrast, the development of ripple-like structures in the other branch demonstrates an apparent asymmetry in domain dynamics during magnetization reversal [1].

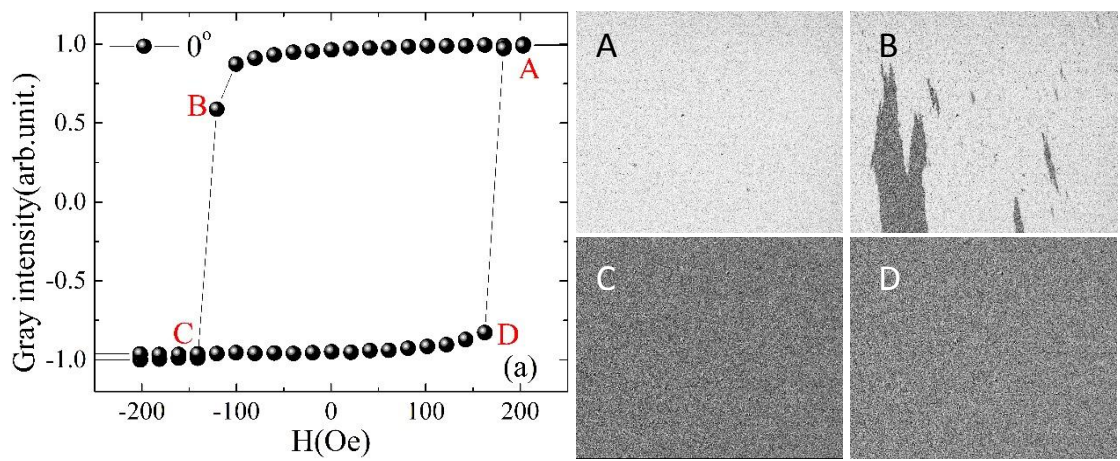

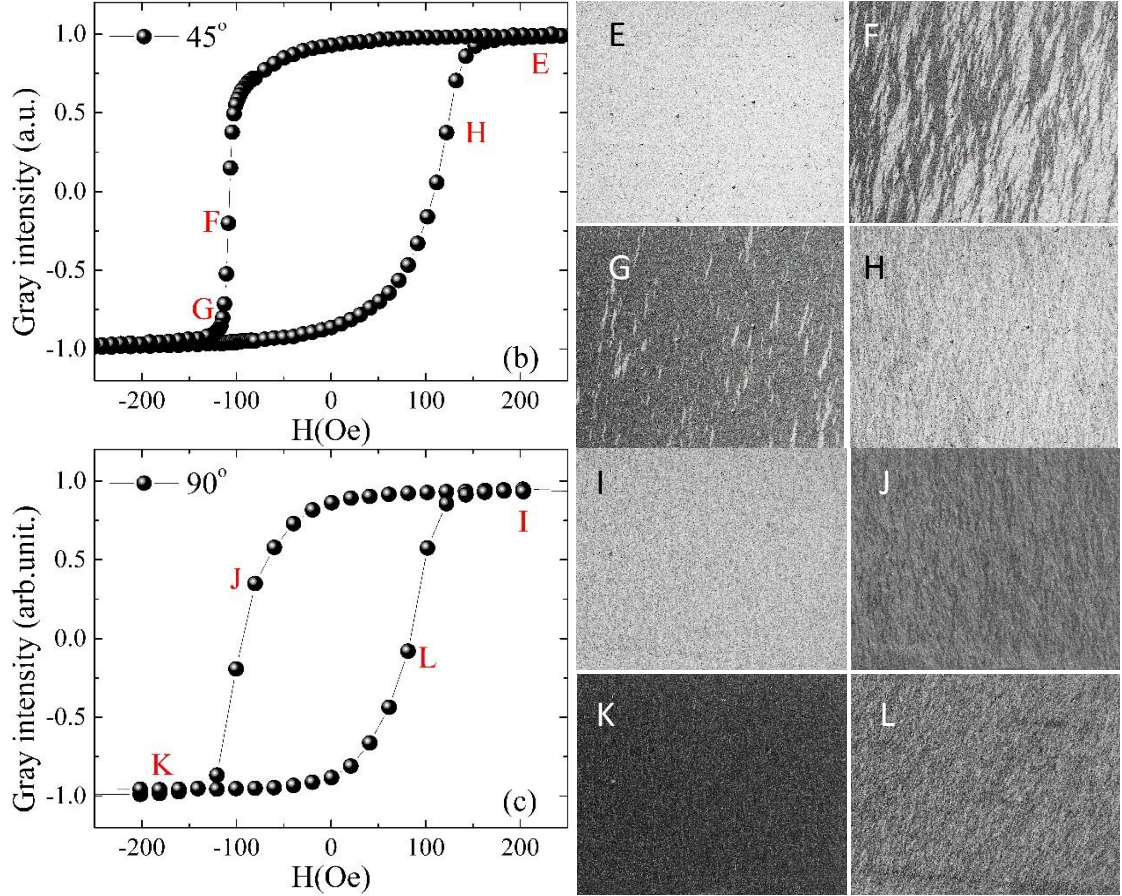

**Fig. 4:** Hysteresis loops of ML with domain images corresponding to the field marked in the hysteresis loop measured along (a) θ=0°, (b) θ=45°, and (c) θ=90° after applying $H_{SAT}$ = 1.5 kOe.

Figure 5(a) shows the hysteresis loops of ML measured as a function of θ after saturating in a 25 kOe magnetic field. It was observed that along θ = 0°, the hysteresis loop remains almost square; however, in contrast to the previous case, now it was accompanied by a shift of 50 Oe along the magnetic field axis, opposite to the $H_{SAT}$ direction. Along θ = 90°, the loop exhibits a rounded symmetric behavior along the magnetic field axis. This shows that the ML exhibits EB, which arises from $K_{UNI}$ introduced due to the large $H_{SAT}$ that aligns the moments of the FePt(L1$_0$) layer in its direction. The polar plot in fig. 5(b) shows the variation of $H_{EB}$ and $H_C$ as a function of θ. It was observed that initially, $H_{EB}$ increases up to θ = 20° and then decreases till θ = 90°. $H_C$ decreases continuously with an increase in θ. As far as the asymmetry in the magnetization reversal is concerned, it was observed that asymmetry was still present for certain range of angles away from the $H_{SAT}$ ($K_{UNI}$) direction; however, the range of asymmetry was reduced to a smaller range with maxima at 20°. This may be attributed to decrease in non-collinearity between the two FeCo layers anisotropies. It may also be noted that the maximum value of $H_{EB}$ is obtained in the direction where asymmetry was maximum.

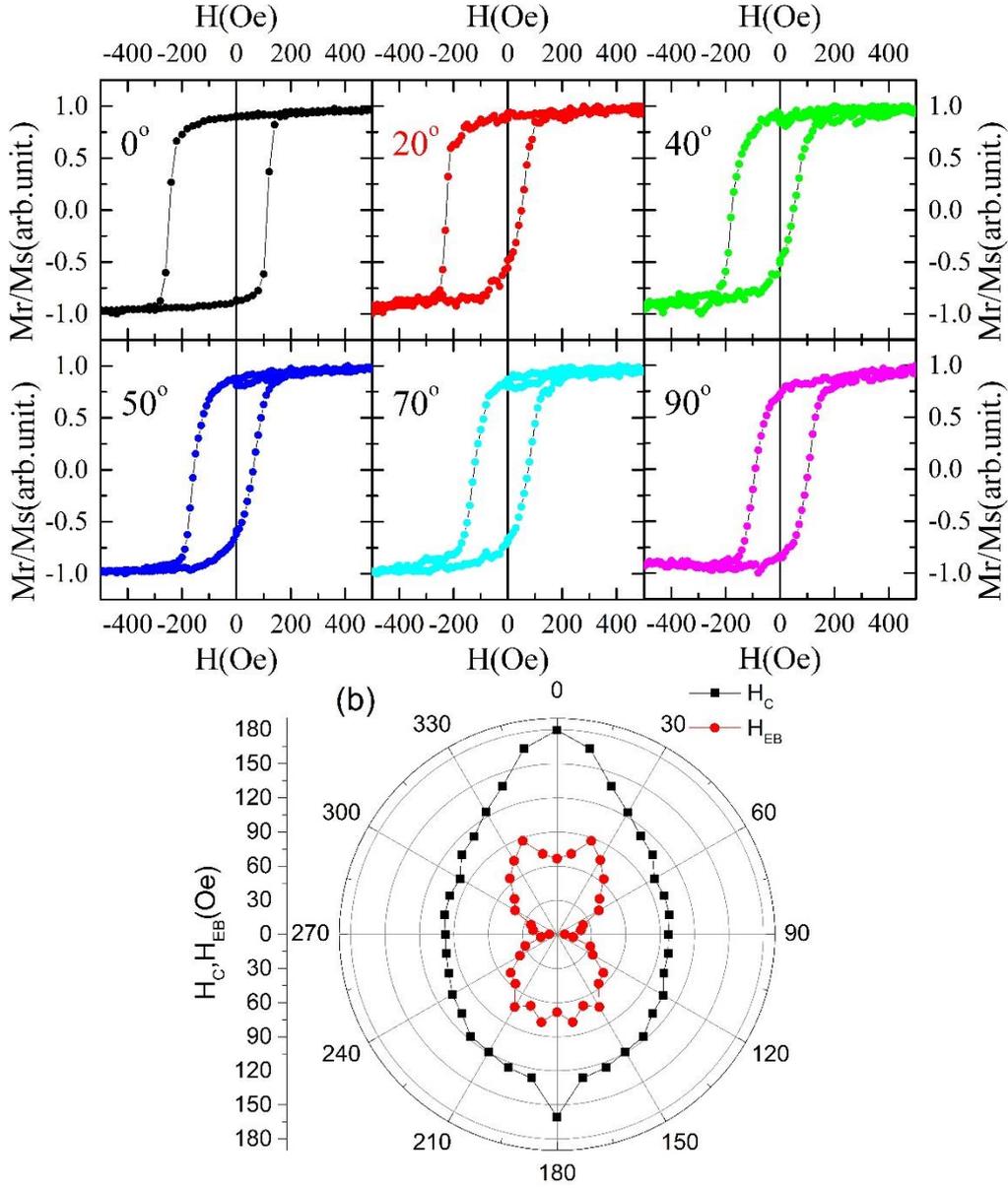

**Fig. 5:** (a) Hysteresis loop of FePt(L1$_0$)/FeCo/CoO/FeCo multilayer measured at various angles after applying H$_{SAT}$ = 25 kOe. (b) Polar plot of variation of H$_C$ and exchange bias field (H$_{EB}$) as a function of θ i.e. angle between H$_{APP}$ and H$_{SAT}$.

Figure 6 shows the hysteresis loop of ML along θ = 0° and 45° after demagnetizing it by applying a 50 kOe magnetic field normal to the film's surface. It was observed that the virgin curve for the first hysteresis loop along θ = 0° starts from almost the center of the magnetization axis, which confirms the demagnetized state of ML. On measuring loop along 45°, the virgin curve shifts upward as the spins are in their remanent state along θ = 0°, and with increasing field, their spins gradually align towards 45°. In addition, loops along 0° and 45° have the same H$_C$ and do not exhibit asymmetric magnetization reversal. This may be because when the film was demagnetized, non-collinearity between FeCo layers disappeared. This further confirms that the asymmetry is only due to non-collinearity between the spins of FeCo layers.

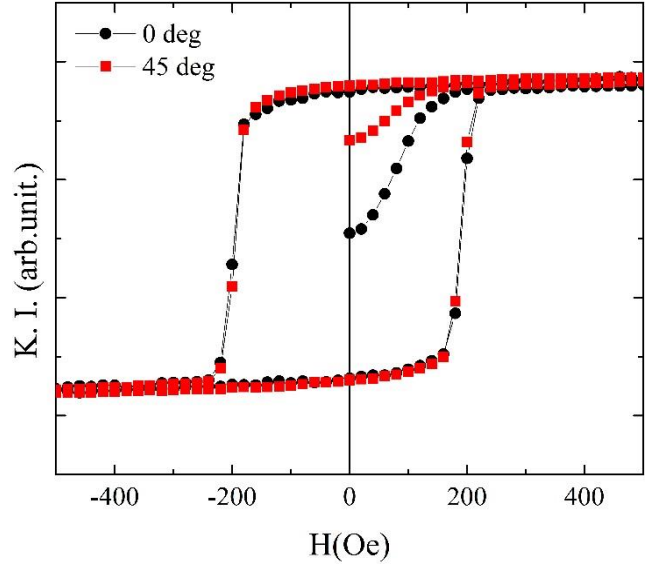

**Fig. 6:** Hysteresis loops of FePt(L1$_0$)/FeCo/CoO/FeCo ML measured at various angles after applying H$_{SAT}$ = 50 kOe field normal to film's surface.

Asymmetric magnetization reversal in magnetic multilayers is generally associated with the presence of EB in the system [1,2,3,4,5]. However, the above results show that the asymmetry in the hysteresis loop can appear even in the absence of EB when the ML was saturated in a 1.5kOe magnetic field. Furthermore, the asymmetry decreased with an increase in saturation field and disappeared when ML was demagnetized. It shows that unidirectional anisotropy may not be solely responsible for asymmetry in EB films. The asymmetry in the present case may be attributed to the non-collinearity between the spins of the dipolar coupled FeCo layers with respect to each other [25,26,27]. The behavior is similar to asymmetrical magnetization reversals reported in Fe/Cr ML [18], Co$_{74}$Tb$_{26}$/Cu/Co$_{88}$Tb$_{12}$ [19] and Hard/Soft SmCo/Fe FM bilayers [28] due to non-collinearity between adjacent magnetic layers.

The non-collinearity between spins of FeCo layers can be explained in terms of the proximity effect through the CoO barrier layer [25,26,27]. Non-collinearity between two FM layers when separated by an AFM layer as in case of Fe/Cr [18], Fe/FeO [25,26], NiFe/FeMn/Co [27], etc. is well reported in literature. The FM layers couple through short-range Heisenberg exchange interaction [18,25,26,27,29], and depending on the thickness of the AFM layer, the coupling can be either FM or AFM. However, in polycrystalline films [29], lateral variation in the thickness of layers results in the appearance of both FM and AFM interactions and the competition between the two leads to non-collinearity between the layers, as shown in fig. 7. The angle of non-collinearity depends on the ratio of the two interactions.

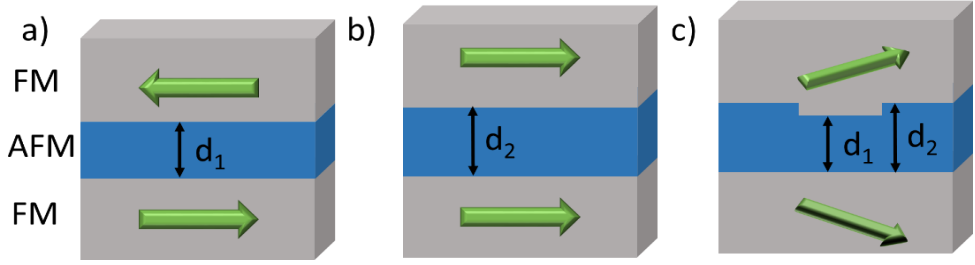

**Fig. 7:** Schematic representation of collinear (a) AFM coupling, (b) FM coupling and (c) non-collinearity due to combination of both FM and AFM coupling due to lateral variation of thickness.

It is to be noted that the present experiments were performed at RT, where CoO, in general, is in a paramagnetic state, which may make the above theory invalid. However, in the present study, CoO is in the form of ultra-thin film (~30 Å) and is in contact with two FM layers at both interfaces. Thus, its Nèèl temperature ($T_N$) may differ from its bulk counterpart (291 K) due to the low thickness and proximity effect that may lift its ordering temperature to 300 K [30,31,32,33]. In addition, the roughness at the CoO interfaces and FeCo/Air interface is large, as observed by XRR, resulting in lateral variation in thickness. This leads to non-collinearity between two FeCo layers giving rise to asymmetry in magnetization reversal.

The initial absence of EB at the low saturation field is attributed to the absence of $K_{UNI}$ in the ML. Firstly, the FePt layer was annealed without an external magnetic field; therefore though a hard magnetic phase (FePt- $L1_0$) with high $H_C$ was obtained, the moments are randomly oriented (demagnetized state), and $H_{SAT}$ ~1.5 kOe is not sufficient to saturate FePt($L1_0$) layer. Secondly, even though CoO is in an AFM state, the AFM ordering may be short-range and random, leading to weak AFM anisotropy, which may not be sufficient to generate $K_{UNI}$. Thus, the absence of $K_{UNI}$, both from the AFM layer and hard FM layer, results in the absence of EB at the low saturation field.

When the ML was saturated in a 25 kOe field, FePt($L1_0$) layer moments will align in the direction of the field. It will then couple with both FeCo layers resulting in $K_{UNI}$ and EB. However, the bottom FeCo layer is in direct contact with the FePt($L1_0$) layer; hence the coupling will be due to interfacial exchange interaction and magneto-static coupling via the stray field. Whereas the top FeCo layer is away from the FePt($L1_0$) layer and separated by the CoO barrier layer, thus the coupling will be only through magnetostatic interaction. Hence, $K_{UNI}$ in the top FeCo will be weak compared to the bottom FeCo layer. Furthermore, the $K_{UNI}$ will try to align the top FeCo layer in the saturation direction of FePt($L1_0$) and compete with the proximity effect, resulting in decease in non-collinearity. Thus, the net angle between magnetization of both FeCo layers will decrease, thereby decreasing the range of angles where asymmetry was observed. Hence, the decrease in the range of the asymmetry region is attributed to the strength of unidirectional anisotropy. Furthermore, the dispersion between the magnetization of the top FeCo layer and the direction of $K_{UNI}$ (in the direction of $H_{SAT}$), gives rise to an asymmetric loop with maximum exchange shift in the direction away from the direction of $K_{UNI}$. This behavior is in accordance with J. Camarero *et al.* [2], where an asymmetric loop is observed away from the field cooling direction in the AFM/FM system due to non-collinearity between $K_U$ of the FM layer and $K_E$ induced by field cooling. They also observed that asymmetry depends on the ratio of the strengths of $K_U$ and $K_E$.

The disappearance of asymmetric magnetization reversal on demagnetizing the ML confirms that the asymmetry disappears with disappearance of non-collinearity between the spins of FeCo layers.

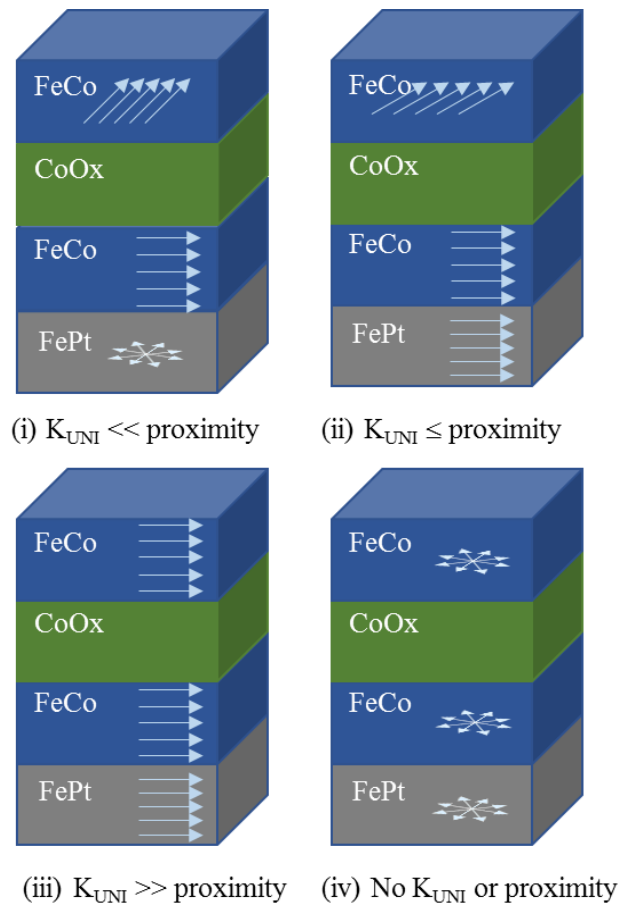

**Fig. 8:** Schematic representation of moments of spins in ML for (i) weak $K_{UNI}$, (ii) moderate $K_{UNI}$ (iii) large $K_{UNI}$ and (iv) absence of proximity and $K_{UNI}$.

Based on the above experiment observation, a competition exists between non-collinearity due to the proximity effect and $K_{UNI}$ due to the hard FePt($L1_0$) layer in FePt($L1_0$)/FeCo/CoO/FeCo ML. The effect of these two interactions on the ML can be summarized in four different cases, as shown in fig. 8: (i) weak $K_{UNI}$ << proximity, (ii) $K_{UNI}$ ≤ proximity, (iii) $K_{UNI}$ >> proximity, and (iv) absence of proximity and $K_{UNI}$.

In the absence of $K_{UNI}$, due to non-collinearity, the ML exhibits asymmetric magnetization reversal similar to the case of 1.5 kOe saturation. Weak or intermediate $K_{UNI}$ gives rise to EB along with asymmetry, as observed in the case of film saturated in 25 kOe. EB will be maximum in the direction where maximum asymmetry is observed. The range of asymmetry will be reduced depending on the strength of $K_{UNI}$ with respect to coupling due to the proximity effect. In the absence of proximity effect and $K_{UNI}$, neither asymmetry nor EB will be observed (similar to the 5 kOe perpendicular field case).

However, another case may arise when $K_{UNI}$ is very large such that FePt(L1$_0$) will align both FeCo layers in the $H_{SAT}$ direction and non-collinearity disappears. It can be achieved by increasing the $H_{SAT}$ or $H_C$ of the FePt(L1$_0$) layer such that the magneto-static coupling due to the stray field increases. This method of the disappearance of non-collinearity due to large $K_{UNI}$ is similar to Fe/FeO ML studied by Diederich *et al.* [25] and Vlasko-Vlasov *et al.* in SmCo/Fe [28], where the application of large external field overcomes the non-collinearity. In this case, ML will exhibit strong EB but without any asymmetry due to large $K_{UNI}$.

Hence, asymmetry irrespective of dipolar or interfacial exchange interaction arises in multilayers due to non-collinearity between two magnetic anisotropies. On the other hand, EB arises due to $K_{UNI}$ induced, either by field cooling in the FM/AFM bilayers or by saturating the hard layer in Hard/Soft FM systems. The explanation, best to our knowledge, justifies all the cases of asymmetry in the literature, with or without EB.

## Conclusions

In the present work, effect of saturation field ($H_{SAT}$) on asymmetric magnetization reversal and EB in FePt(L1$_0$)/FeCo/CoO/FeCo ML structure has been studied as a function of varying azimuthal angle. For low $H_{SAT}$, asymmetric magnetization reversal takes place for a large range of angles without any EB, whereas EB with asymmetric magnetization reversal for a reduced angular range was observed for high $H_{SAT}$. Asymmetry is attributed to non-collinearity between magnetic anisotropy axis of two FeCo layer through barrier CoO layer whereas EB arises as a result of unidirectional anisotropy $K_{UNI}$ induced by FePt(L1$_0$) layer due to $H_{SAT}$ which pins the FeCo layer. Asymmetry range decreases due to decrease in non-coliinear between the two FeCo layers as a result of coupling by FePt(L1$_0$) layer. It is also proposed that further increase in $K_{UNI}$ of FePt(L1$_0$) layer will lead to EB without any asymmetric reversal due to loss of non-collinearity. Furthermore, based on the experimental observation, a justification is provided to correlate origin of asymmetry and EB in interlayer coupled EB system.

## Acknowledgement:

We acknowledge Dr. Zaineb Hussain and Er. Anil Kumar, UGC-DAE CSR, Indore, India, for Kerr microscopy and XRR measurements respectively.